\documentclass[reprint,twocolumn]{revtex4}%
\usepackage{amsfonts}
\usepackage{amsmath}
\usepackage{amssymb}
\usepackage{charter}
\usepackage{graphicx}%
\providecommand{\U}[1]{\protect \rule{.1in}{.1in}}
\begin{document}
\title{Nonclassical properties of Hermite polynomial's excitation on squeezed vacuum
and its decoherence in phase-sensitive reservoirs}
\author{{\small Shi-You Liu}$^{1}${\small , Ya-Zhou Li}$^{1}${\small ,Li-Yun
Hu}$^{1,2\ast}$\thanks{Corresponding author. Email: hlyun2008@126.com}%
,{\small Jie-Hui Huang}$^{1,2}${\small , Xue-Xiang Xu}$^{1}${\small ,}
{\small and Xiang-Yang Tao}$^{1}$}
\affiliation{$^{1}${\small Center for Quantum Science and Technology, Jiangxi Normal
University, Nanchang 330022, China}}
\affiliation{$^{2}${\small Beijing Computational Science Research Center, Beijing, 100084,
China}}

\begin{abstract}
{\small We introduce Hermite polynomial excitation squeezed vacuum (SV)
\ }$H_{n}(\hat{O})S\left(  r\right)  \left \vert 0\right \rangle $ {\small with
}$\hat{O}=\mu a+\nu a^{\dagger}${\small . We investigate analytically the
nonclassical properties according to Mandel's Q parameter, second correlation
function, squeezing effect and the negativity of Wigner function (WF). It is
found that all these nonclassicalities can be enhanced by }$H_{n}(\hat{O}%
)${\small operation and adjustable parameters }$\mu${\small and }$\nu
${\small . In particular, the optimal negative volume }$\delta_{opt}%
${\small of WF can be achieved by modulating }$\mu${\small and }$\nu$
{\small for }$n\geqslant2,${\small while }$\delta${\small \ is kept unchanged
for }$n=1${\small .}$\ ${\small Furthermore, the decoherence effect of
phase-sensitive enviornment on this state is examined. It is shown that
}$\delta$ {\small with bigger }$n${\small diminishes more quickly than that
with lower }$n${\small , which indicates that single-photon subtraction SV
presents more roboustness. Parameter }$M${\small of reservoirs can be
effectively used to improve the nonclassicality.}



\end{abstract}
\maketitle

\section{Introduction}

Nonclassical light fields play a critical rule in quantum optics and quantum
information process \cite{1}. Generation and manipulation of these states have
attracted much attention to obtain a more effective quantum processing, such
as teleportation, dense coding and quantum cloning. There are many schemes
proposed to realize this purpose. Non-Gaussian operation, say photon addition
and photon subtraction, has been widely employed to enhance the nonclassical
properties of the input states \cite{2,3,4,5,6,6a,6b,6c,6d,6e,6f,6g,6h}. For
instance, a quantum-to-classical transition has been realized experimentally
through single-photon--added coherent states of light \cite{7}. For any
photon--addition coherent state in the dissipative channel, its nonclassical
properties are examined theoretically \cite{8} by the analytical expression of
the Wigner function (a Laguerre--Gaussian function). As another example,
photon subtraction or addition has been used to improve entanglement between
Gaussian states and the average fidelity of quantum teleportation \cite{9,10}.

On the other hand, superposition of operators, such as $a^{\dagger}a$,
$ta^{\dagger}+ra$, $a^{2}+b^{2}$, $a^{\dagger2}+b^{\dagger2}$, are also
applied to generate nonclassical states \cite{11,12,13,13a,13b,14}. For
example, the $ta^{\dagger}+ra$ operator is uesed to realize quantum state
engineering and improve quantum entanglement or non-Gaussian entanglement
distillation or the effect of quantum teleportation \cite{11,12}. In addition,
this superposition operation is employed to enhance the degree of entanglement
of even entangled coherent state \cite{15}. Thus it will be interesting to
investigate the different combination of elementary non-Gaussian operations to
manipulate nonclassical quantum states. As a kind of polynomials states, for
instace, the squeezed Hermoite states is found to be the minimum uncertain
states for amplitude-squared squeezing \cite{16}. In addition, the squeezed
two-variable Hermite polynomial state is shown to be the minimum uncertain
states for amplitude-squared squeezing, which is called as the sum-frequency
squeezing states \cite{17}.

Recently, the Hermite polynomial's coherent state $H_{n}\left(  Q\right)
\left \vert \alpha \right \rangle $ is introduced \cite{18}, where
$Q=(a+a^{\dagger})/\sqrt{2}$ is the coordinate operator and $\left \vert
\alpha \right \rangle $=$\exp \{ \alpha a^{\dagger}-\alpha^{\ast}a\} \left \vert
0\right \rangle $ is the Glauber coherent state. Then some nonclassical
properties are discussed in details. In this paper, we shall introduce another
kind of non-Gaussian state, which can be generated by operating Hermite
polynomial of superposition of coherent photon-subtraction and addition (HPS),
i.e., $H_{n}\left(  \mu a+\nu a^{\dagger}\right)  $ on single-mode squeezed
vacuum (SV) $S\left(  r\right)  \left \vert 0\right \rangle $. Single photon
subtraction/addition SV, Hermite polynomial's subtraction $H_{n}\left(  \mu
a\right)  $ and addition $H_{n}\left(  \nu a^{\dagger}\right)  $ SV can be
considered as special cases of the HPS. It is interesting to notice that the
HPS-SV can be generated by superposing some photon-addition and
photon-subtraction SVs. As far as we know, there is no report in literature before.

This paper is arranged as follows. In section 2, we shall derive the
normalization factor $N_{\mu_{1},\nu_{1}}$ for the non-Gaussian states. It is
shown that $N_{\mu_{1},\nu_{1}}$ is just the Legendra polynomials, which is
needed for clearly discussing the statistical properties of the HPS-SV. In
section 3, we shall discuss nonclassical properties of the HPS-VS by
ananlytically deriving Mandel's Q parameter, second correlation function,
photon-number distribution, squeezing effect. In section 4, the Wigner
function (WF) of the HPS-SV is obtained by using the property of\ Weyl ordered
operators' invariance under similar transformations. In particular, the
nonclassical property is presented according to the negativity of the WF.
Section 5 is devoted to considering the effect of phase-sensitive reservoirs
on the HPS-SV in terms of the negativity of WF. The last section is used to
draw a conclusion.

\section{The HPS-SV and its normalization}

The HPS-SV can be generated by operating Hemite polynomial operator
$H_{n}\left(  \mu a+\nu a^{\dagger}\right)  $ on single-mode squeezed vacuum
$S\left(  r\right)  \left \vert 0\right \rangle $,
\begin{equation}
\left \vert \Psi \right \rangle _{H}=N_{\mu_{1},\nu_{1}}H_{n}(\hat{O})S\left(
r\right)  \left \vert 0\right \rangle ,\hat{O}\equiv \mu a+\nu a^{\dagger},
\label{h1}%
\end{equation}
where $N_{\mu_{1},\nu_{1}}$ is the normailzarion factor to be determined, and
$S\left(  r\right)  =\exp \{ \frac{r}{2}\left(  a^{2}-a^{\dag2}\right)  \}$ is
the squeezing operator with $r$ being squeezing parameter, and $H_{n}(x)$ is
the single-variable Hermite polynomials. $a$ and $a^{\dagger}$ are the Bose
anahilate\ and creation operator, respectively, satisfying communicative
relation $[a,a^{\dagger}]=1$.

In order to calculate $N_{\mu_{1},\nu_{1}}$, using the transformation relation
of single-mode squeezed operator \cite{19}, $S^{\dagger}\left(  r\right)
aS\left(  r\right)  =a\cosh r-a^{\dag}\sinh r$, $S^{\dagger}\left(  r\right)
a^{\dag}S\left(  r\right)  =a^{\dag}\cosh r-a\sinh r$, we can get
\begin{equation}
S^{\dagger}\left(  r\right)  H_{n}(\hat{O})S\left(  r\right)  =H_{n}(\hat
{O}_{1}),(\hat{O}_{1}\equiv \mu_{1}a+\nu_{1}a^{\dagger}), \label{h3}%
\end{equation}
where $\mu_{1}=\mu \cosh r-\allowbreak \nu \sinh r$, $\nu_{1}=\nu \cosh r-\mu \sinh
r.$ Thus the factor $N_{\mu_{1},\nu_{1}}$ can be calculated according to the
normalization $1=\left \langle \Psi \right \vert \left.  \Psi \right \rangle _{H},$
i.e.,
\begin{equation}
N_{\mu_{1},\nu_{1}}^{-2}=\left \langle 0\right \vert H_{n}(\hat{O}_{1}^{\dagger
})H_{n}(\hat{O}_{1})\left \vert 0\right \rangle . \label{h5a}%
\end{equation}
Then further employing the generating of function of single-variable Hermite
polynomial,
\begin{equation}
H_{n}\left(  x\right)  =\left.  \frac{\partial^{n}}{\partial t^{n}}%
e^{-t^{2}+2tx}\right \vert _{t=0}, \label{h6}%
\end{equation}
and the following operator identity \cite{19} $e^{A+B}=e^{A}e^{B}e^{-\frac
{1}{2}\left[  A,B\right]  }=e^{B}e^{A}e^{\frac{1}{2}\left[  A,B\right]  }$,
which is valid for $[A,[A,B]]=[B,[A,B]]=0$, we can put Eq.(\ref{h5a}) into the
form%
\begin{align}
N_{\mu_{1},\nu_{1}}^{-2}  &  =\left.  \frac{\partial^{2n}}{\partial \tau
^{n}\partial t^{n}}\exp \left \{  -A\left(  \tau^{2}+t^{2}\right)  +4\nu_{1}%
^{2}t\tau \right \}  \right \vert _{\tau,t=0}\nonumber \\
&  =2^{n}n!B^{n}P_{n}\left(  2\nu_{1}^{2}/B\right)  , \label{h8}%
\end{align}
where $A=1-2\mu_{1}\nu_{1},$ $B=\sqrt{4\nu_{1}^{4}-A^{2}}$, $P_{n}$ is the
Legendre polynomial, and we have used the new formula \cite{20}
\begin{align}
&  \frac{\partial^{2m}}{\partial t^{m}\partial \tau^{m}}\left.  \exp \left(
-t^{2}-\tau^{2}+\frac{2x\tau t}{\sqrt{x^{2}-1}}\right)  \right \vert
_{t,\tau=0}\nonumber \\
&  =\frac{2^{m}m!}{\left(  x^{2}-1\right)  ^{m/2}}P_{m}\left(  x\right)  ,
\label{h9}%
\end{align}
in the last step of Eq.(\ref{h8}). Eq.(\ref{h8}) is just the analytical
expression of the normalization factor $N_{\mu_{1},\nu_{1}}^{-2}$.

In particular, when $(\mu,\nu)=(1,0),\allowbreak(0,1),$ leading to $(\mu
_{1},\nu_{1})=(\cosh r,-\sinh r)$,\allowbreak \ $(-\allowbreak \sinh r,\cosh
r)$, and $B=\sqrt{1-2e^{2r}}$, $\sqrt{1+2e^{-2r}}$, thus Eq.(\ref{h8}) reduces
to
\begin{align}
N_{1,0}^{-2}  &  =2^{n}n!\left(  \sqrt{1-2e^{2r}}\right)  ^{n}P_{n}\left(
\frac{2\sinh^{2}r}{\sqrt{1-2e^{2r}}}\right)  ,\label{h10}\\
N_{0,1}^{-2}  &  =2^{n}n!\left(  \sqrt{1+2e^{-2r}}\right)  ^{n}P_{n}\left(
\frac{2\cosh^{2}r}{\sqrt{1+2e^{-2r}}}\right)  , \label{h11}%
\end{align}
which are just the normalization factors of Hermite subtraction and Hermite
addition squeezed vacuum, respectively. In addition, when the squeezing
parameter $r=0\ $leading to $\mu_{1}=\mu$, $\nu_{1}=\nu,$ i.e., the HPS vacuum
($\left \vert \Psi \right \rangle _{HPS}\rightarrow \left \vert \Psi \right \rangle
$), we see
\begin{equation}
\left \vert \Psi \right \rangle \equiv N_{\mu,\nu}H_{n}(\hat{O})\left \vert
0\right \rangle ,\text{ }N_{\mu,\nu}^{-2}=\left \{  N_{\mu_{1},\nu_{1}}%
^{-2}\right \}  _{\left(  \mu_{1},\nu_{1}\right)  \rightarrow \left(  \mu
,\nu \right)  }. \label{h11a}%
\end{equation}

In the state, we can get the average%
\begin{equation}
\left \langle a^{\dag l}a^{k}\right \rangle =\frac{N_{\mu,\nu}^{2}\nu
^{2n}\left(  2\lambda \right)  ^{l+k}\left(  n!\right)  ^{2}}{\lambda
^{2n}\left(  n-l\right)  !\left(  n-k\right)  !}F_{n-l,n-k}\left(  \lambda
^{2}\right)  , \label{h11b}%
\end{equation}
where $\lambda=\nu/\sqrt{1-2\mu \nu}$ and we have define a special function
whose mother function is given by%
\begin{equation}
F_{m,n}\left(  \lambda^{2}\right)  \equiv \left.  \frac{\partial^{m}}{\partial
s^{m}}\frac{\partial^{n}}{\partial t^{n}}e^{-t^{2}-s^{2}+4st\lambda^{2}%
}\right \vert _{s=t=0}. \label{h14b}%
\end{equation}
From Eq.(\ref{h11b}) one can see that $\left \langle a^{\dag l}a^{k}%
\right \rangle =\left \langle a^{\dag k}a^{l}\right \rangle ^{\ast}$,
$\left \langle a^{\dag l}\right \rangle =\left \langle a^{l}\right \rangle ^{\ast
}$ and $\left \langle a\right \rangle =0$, as expected. In particular, when
$k=l,$ Eq.(\ref{h11b}) reduces to
\begin{align}
\left \langle a^{\dag l}a^{l}\right \rangle  &  =\frac{2^{n+l}N_{\mu,\nu}%
^{2}\left(  n!\right)  ^{2}}{\left(  n-l\right)  !}v^{2l}K^{(n-l)/2}%
P_{n-l}\left(  \frac{2\nu^{2}}{\sqrt{K}}\right) \label{11c}\\
(K  &  =4\nu^{4}-\left(  1-2\mu \nu \right)  ^{2},n\geqslant l),\nonumber
\end{align}
Eq.(\ref{h11b}) shall be useful for further calculations.

\section{Nonclassical properties of the HPS-VS}

In this section, we study the nonclassical properties of the HPS-VS according
to Mandel's Q parameter, photon-number distribution, and squeezing effect.

\subsection{Mandel's Q parameter and second-order correlation function}

We examine the sub-Poissonian photon statistics by using Mandel's Q parameter
\cite{21}, which is defined as%
\begin{equation}
Q_{M}=\frac{\left \langle a^{\dag2}a^{2}\right \rangle _{H}}{\left \langle
a^{\dag}a\right \rangle _{H}}-\left \langle a^{\dag}a\right \rangle _{H}.
\label{h12}%
\end{equation}
Super-Poissonian, Poissonian, and sub-Poissonian statistics correspond to
$Q_{M}>0$, $Q_{M}=0,$ and $Q_{M}<0,$respectively. In order to obtain the
result (\ref{h12}), it will be convinient to derive some average values:
$\left \langle a^{2}\right \rangle $, $\left \langle a^{4}\right \rangle ,$
$\left \langle a^{\dag}a\right \rangle ,$ $\left \langle a^{\dag3}a\right \rangle
$\ and $\left \langle a^{\dag2}a^{2}\right \rangle $ under the state $\left \vert
\Psi \right \rangle $ (\ref{h11a}). These averages are obtained from
Eq.(\ref{h11b}). Thus under the state $\left \vert \Psi \right \rangle _{H}$,
these corresponding average values ($\left \langle ...\right \rangle
_{H}=\left \{  \left \langle S^{\dag}...S\right \rangle \right \}  _{\mu
\rightarrow \mu_{1},\nu \rightarrow \nu_{1}}$) are given by%
\begin{align}
\left \langle a^{\dag}a\right \rangle _{H}  &  =\left \{  \left \langle a^{\dag
}a\right \rangle \cosh2r+\sinh^{2}r\right. \nonumber \\
&  \left.  -\frac{\sinh2r}{2}\left \langle a^{\dag2}+a^{2}\right \rangle
\right \}  _{\mu \rightarrow \mu_{1},\nu \rightarrow \nu_{1}}, \label{h16}%
\end{align}
and%
\begin{align}
&  \left \langle a^{\dag2}a^{2}\right \rangle _{H}\nonumber \\
&  =\left \{  \frac{1}{4}\left(  3\cosh4r+1\right)  \left \langle a^{\dag2}%
a^{2}\right \rangle +\frac{1}{4}\left \langle a^{\dag4}+a^{4}\right \rangle
\sinh^{2}2r\right. \nonumber \\
&  +\left(  \sinh2r-\frac{3}{4}\sinh4r\right)  \left \langle a^{\dag2}%
+a^{2}\right \rangle \allowbreak \nonumber \\
&  -\frac{1}{2}\left \langle a^{\dag}a^{3}+a^{\dag3}a\right \rangle
\sinh4r+4\left \langle a^{\dag}a\right \rangle \left(  \allowbreak3\cosh
^{2}r-1\right)  \sinh^{2}r\nonumber \\
&  \left.  +(3\cosh^{2}r-2)\sinh^{2}r\right \}  _{\mu,\nu \rightarrow \mu_{1}%
,\nu_{1}}. \label{h15}%
\end{align}
Substituting Eq.(\ref{h11b}) into Eq.(\ref{h12}), we can get the Mandel's Q
parameter. The second-order correlation function \cite{22} $g^{(2)}%
=\left \langle a^{\dag2}a^{2}\right \rangle _{H}/\left \langle a^{\dag
}a\right \rangle _{H}^{2}$ can also be gotten by using Eqs.(\ref{h16}) and
(\ref{h17}). In particular, when $n=0$ (corresponding to the squeezed vacuum),
the Mandel's Q parameter and the second-order correlation function are given
by $Q_{M}=\cosh2r>1$, and$\;g^{(2)}=3+1/\sinh^{2}r>3$, respectively.

In order to clearly see the effects of Hermite polynomial on squeezed vacuum,
the numerical calculation results of squeezing parameter $r$ and Mandel's Q
parameter $Q_{M}$, $g^{(2)}$ are plotted as the functions of squeezing
parameter $r$ in Figs.1, 2, respectively. From Fig.1(a) for a given $(\mu
,\nu)$=$(1,1)$, it is easy to see that the HPS-SV presents a sub-Poissonian
statistics (except for $n=0$) in a small region of parameter $r\lesssim0.5$
and the value of $Q_{M}$ increases with $r$. However, the absolute value of
$Q_{M}$ decreases with $n$ in this region. For a given value of $n=2$, and
several asymmetrical cases of $(\mu,\nu),$ on one hand, from Fig.1(b) one can
see that the negative feature can be enhanced by Hermite polymials addition
$H_{2}(a^{\dagger})$ operation rather than Hermite polymials subtraction
operator $H_{2}(a)$. The latter shows a similar trend to the SV due to their
similar photon-number distributions. On the other hand, the asymmetrical
coherent superposition of subtraction and addition (say, $H_{2}(a+9a^{\dagger
})$) can be more effective for improving the negative feature of $Q_{M}$ than
$H_{2}(a)$ and $H_{2}(a+a^{\dagger})$. In addition, from Fig.2 one can get
similar results for the second-order correlation function. For instance, the
HPS-SV appears antibunching effect in a small region due to the Hermite
operation (except for $n=0$).

\begin{figure}[ptb]
\label{Fig1} \centering \includegraphics[width=6cm]{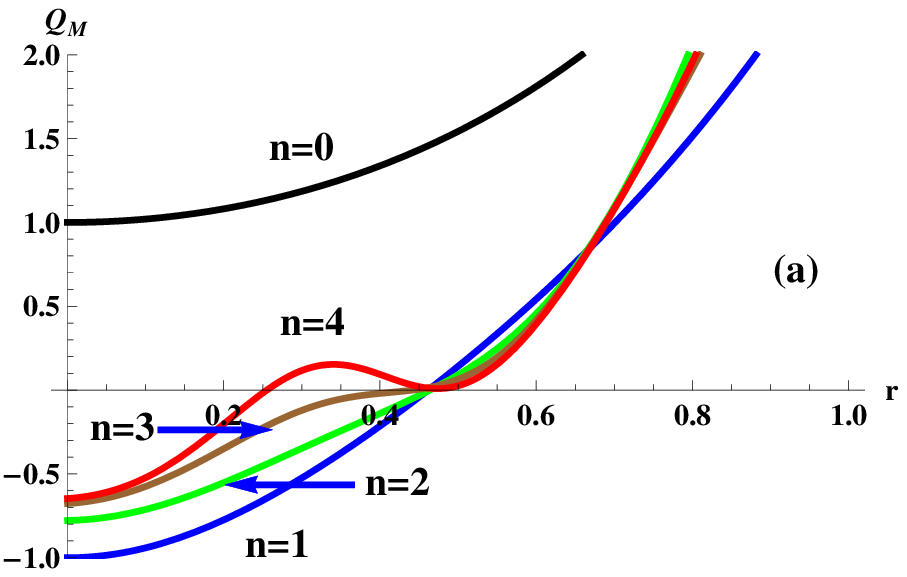} \newline%
\includegraphics[width=6cm]{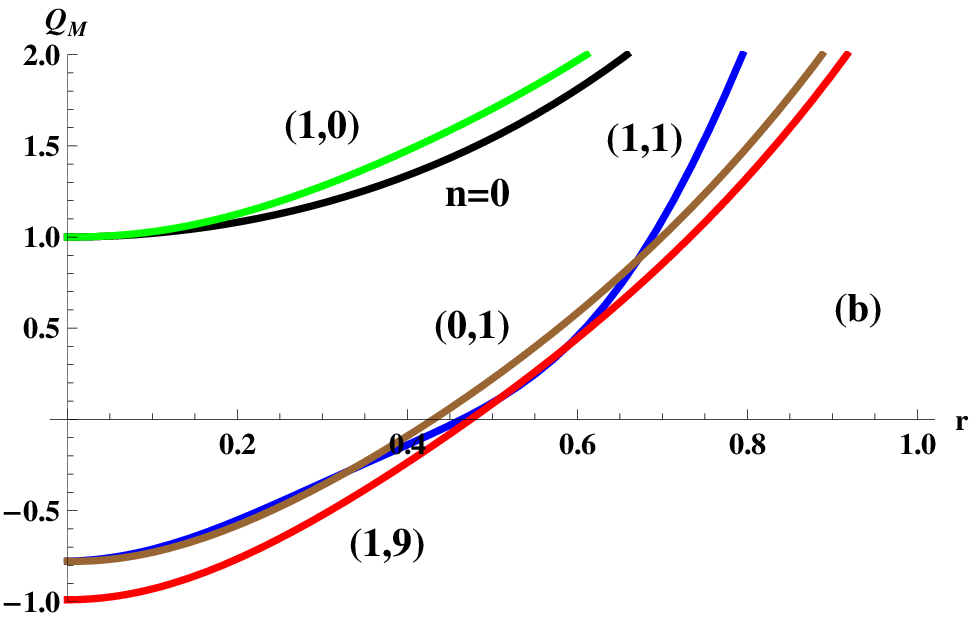}\caption{{}(Color online) Mandel`s $Q$
parameter $Q_{M}$ as a function of squeezing parameter $r$ for several
different values of $n$ and ($\mu,\nu$). (a) ($\mu,\nu$)$=$($1,1$); (b)
$n=2.$}%
\end{figure}

\begin{figure}[ptb]
\label{Fig2} \centering \includegraphics[width=6cm]{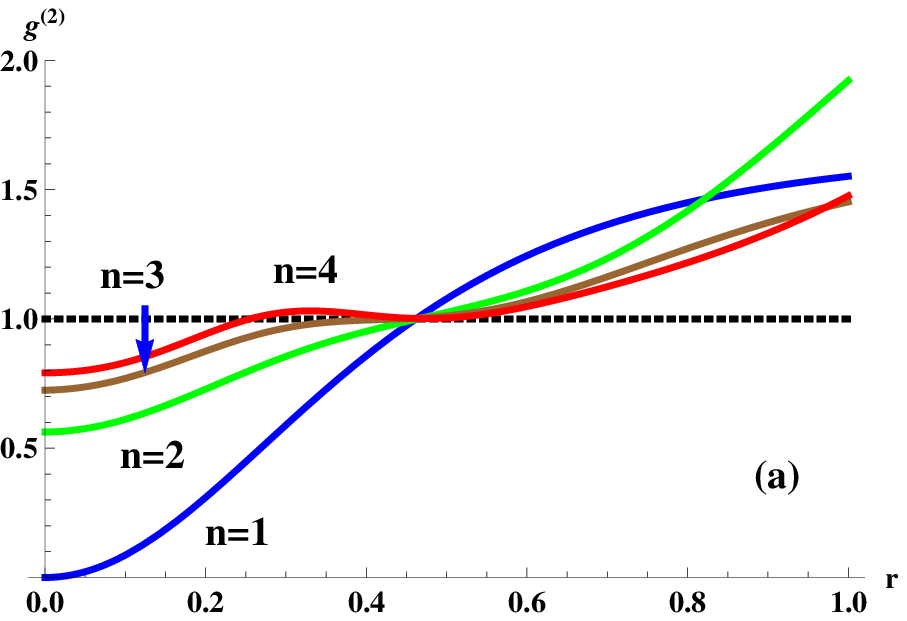} \newline%
\includegraphics[width=6cm]{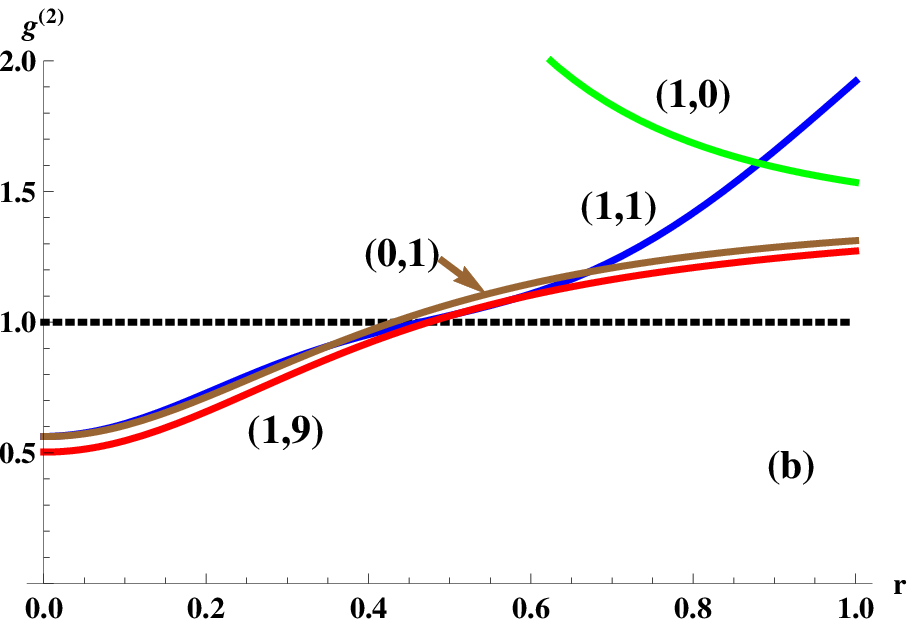}\  \caption{{}(Color online) The
second-order correlation function $g^{(2)}$ as the function of squeezing
parameter $r$ for several different values of $n$ and ($\mu,\nu$). (a)
($\mu,\nu$)$=$($1,1$); (b) $n=2.$}%
\end{figure}

\subsection{Photon-number distribution}

Now, we discuss the photon-number distribution of the HPS-SV. In this field,
the photon-number distribution (PND) of finding $m$ photons is given by
$P_{H}=\left \vert N_{\mu_{1},\nu_{1}}\left \langle m\right \vert H_{n}(\hat
{O})S\left(  r\right)  \left \vert 0\right \rangle \right \vert ^{2}$. Employing
the un-normalized coherent state $\left \vert \alpha \right \rangle =\exp[\alpha
a^{\dag}]\left \vert 0\right \rangle $ ($\left \langle 0\right.  \left \vert
\alpha \right \rangle =1$) \cite{24,25}, leading to $\left \vert m\right \rangle
=\frac{1}{\sqrt{m!}}\frac{\mathtt{d}^{m}}{\mathtt{d}\alpha^{m}}\left \vert
\alpha \right \rangle \left \vert _{\alpha=0}\right.  ,$ $\left(  \left \langle
\beta \right.  \left \vert \alpha \right \rangle =e^{\alpha \beta^{\ast}}\right)
$, and the SV \cite{19,22}%
\begin{equation}
S\left(  r\right)  \left \vert 0\right \rangle =\text{sech}^{1/2}r\exp \left(
-\frac{1}{2}a^{\dag2}\tanh r\right)  \left \vert 0\right \rangle , \label{h17}%
\end{equation}
as well as Eq.(\ref{h6}) ($e^{\alpha a}a^{\dag}e^{-\alpha a}=a^{\dag}+\alpha
$), we have%
\begin{align}
&  \left \langle m\right \vert H_{n}(\hat{O})S\left(  r\right)  \left \vert
0\right \rangle \nonumber \\
&  =\frac{\text{sech}^{1/2}r}{\sqrt{m!}}\left.  \frac{\partial^{m}\partial
^{n}}{\partial \alpha^{\ast m}\partial t^{n}}e^{-A_{1}^{2}t^{2}-B_{1}^{2}%
\alpha^{\ast2}+2t\alpha^{\ast}C_{1}}\right \vert _{t=\alpha^{\ast}=0},
\label{h18}%
\end{align}
where we have set $A_{1}^{2}=1+2\mu^{2}\tanh r-2\mu \nu$, $B_{1}^{2}=\frac
{1}{2}\tanh r$, $C_{1}=\nu-\mu \tanh r.$ Thus the photon-number distribution
is
\begin{align}
P_{H}\left(  m\right)   &  =N_{\mu_{1},\nu_{1}}^{2}m!\left(  n!\right)
^{2}\text{sech}r\nonumber \\
&  \times \left \vert \left.  \frac{\partial^{m}\partial^{n}}{\partial
\alpha^{\ast m}\partial t^{n}}e^{-A_{1}^{2}t^{2}-B_{1}^{2}\alpha^{\ast
2}+2t\alpha^{\ast}C_{1}}\right \vert _{t=\alpha^{\ast}=0}\right \vert ^{2},
\label{h20}%
\end{align}
which is the PND of the HPS-SV. It is easy to see that Eq.(\ref{h20}) just
reduces to the photon-number distribution of SV when $n=0$; while for $r=0$,
(i.e., the Hermite polynomial vacuum state $H_{n}(\hat{O})\left \vert
0\right \rangle $), the PND is given by
\begin{align}
P_{H}\left(  m\right)   &  =N_{\mu,\nu}^{2}m!\left(  n!\right)  ^{2}%
\nonumber \\
&  \times \left \vert \sum_{l=0}^{[n/2]}\frac{\left(  2\mu \nu-1\right)
^{l}\left(  2\nu \right)  ^{n-2l}}{l!\left(  n-2l\right)  !}\delta
_{m,n-2l}\right \vert ^{2}.
\end{align}
In particular, when $m=n$, using the formula \cite{20}
\begin{equation}
P_{m}\left(  x\right)  =x^{m}\sum_{l=0}^{[m/2]}\frac{m!}{2^{2l}l!l!(m-2l)!}%
\left(  1-\frac{1}{x^{2}}\right)  ^{l}, \label{h21}%
\end{equation}
we have
\begin{align}
P_{H}\left(  n\right)   &  =N_{\mu,\nu}^{2}\left(  n!\right)  ^{2}%
2^{2n}\text{sech}r\left \vert \text{ }D_{1}^{n/2}P_{n}\left(  C_{1}/\sqrt
{D_{1}}\right)  \right \vert ^{2},\label{h23}\\
&  \left(  D_{1}=C_{1}^{2}-A_{1}^{2}B_{1}^{2}\right)  .\nonumber
\end{align}

In Fig.3, the PND is poltted for different values $\left(  \mu,\nu \right)  $,
$r$ and $n$, from which one can see that (i) by modulating the order of
Hermite polynomials, one has able to change the position of peak [see Fig.3
(a) and (d)]; (ii) for a small squeezing (say $r=0.3$), the peak of PND is
mainly located at $n$ [see Fig. 3 (a), (b), (d)]; (iii) for a large squeezing
(say $r=0.9$), the peak moves to the small photon-number region (see
Fig.3(b)); (iv) in addition, the PND can also be modulated by the parameters
[see Fig.3 (a) and (c)], expecially for $n\geqslant2$.

\begin{figure}[ptb]
\label{Fig4} \centering \includegraphics[width=4cm]{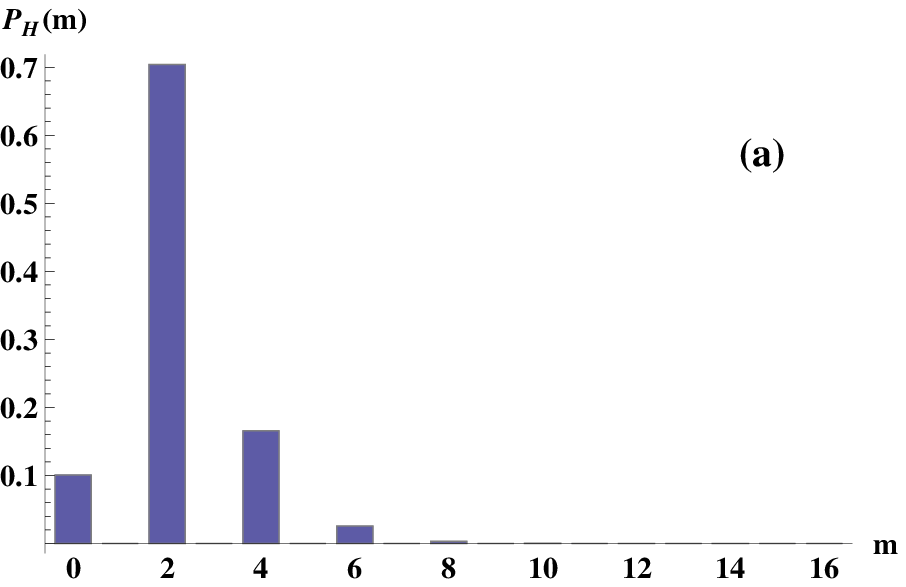}
\  \  \includegraphics[width=4cm]{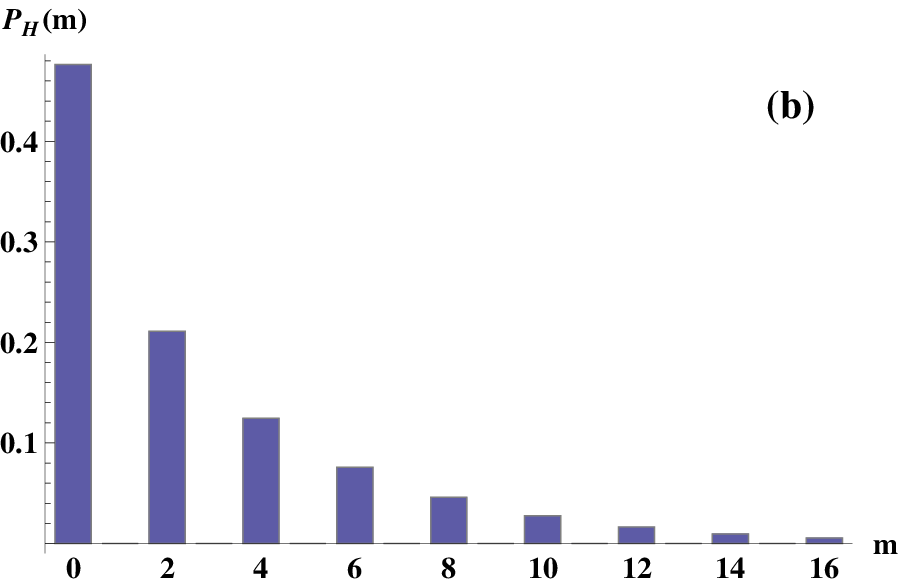}\  \  \newline%
\includegraphics[width=4cm]{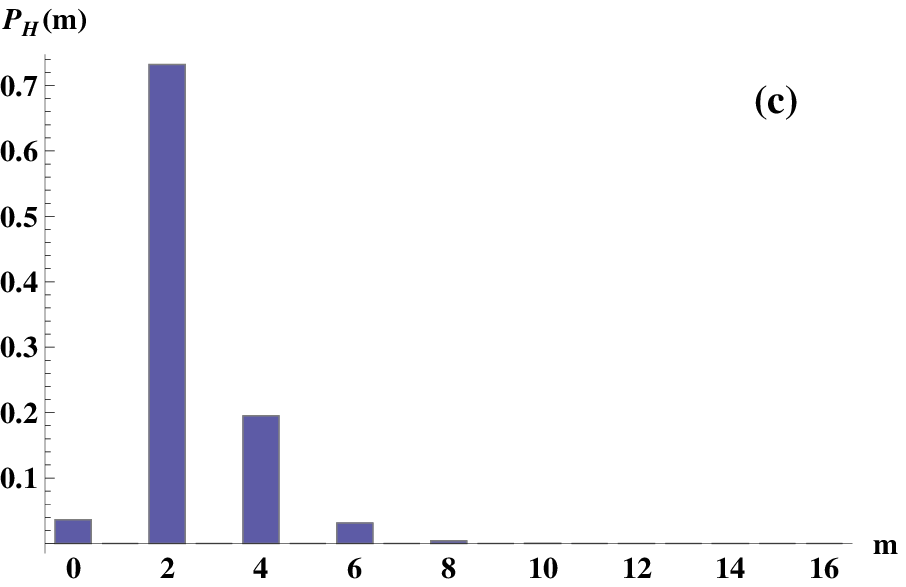}
\includegraphics[width=4cm]{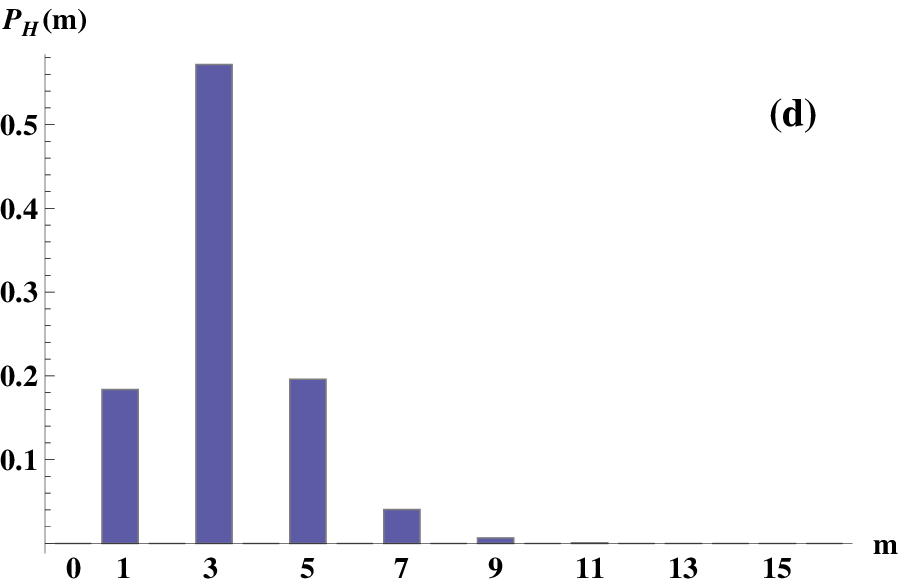}\  \caption{{}(Color online) The
photon-number of the HPS-SV distribution as a function of $m$ for several
different parameters $r$, $n$ and ($\mu,\nu$). (a) $n=2,r=0.3$, $\mu
,\nu=1/\sqrt{2}$; (b) $n=2,r=0.9$, $\mu,\nu=1$/$\sqrt{2}$; (c) $n=2,r=0.3$,
$\mu=1,\nu=3$; (d) $n=3,r=0.3,$ $\mu,\nu=1$/$\sqrt{2}$.}%
\end{figure}

\subsection{Squeezing effects}

In this subsection, we consider the squeezing effects of the HPS-SV,
eapecially from the Hermite polynomial operation. First, we examine the wave
function which can reflect the the squeezing effect of quantum state to some
extent. Using the natural expression of single-mode squeezing operator in the
momentum representation $\left \vert p\right \rangle $ \cite{27},
\begin{equation}
S\left(  r\right)  =\sqrt{u}\int_{-\infty}^{\infty}dp\left \vert
up\right \rangle \left \langle p\right \vert ,u=e^{r}, \label{h24}%
\end{equation}
which leads to $S^{\dag}\left(  r\right)  \left \vert p\right \rangle
=1/\sqrt{u}\left \vert p/u\right \rangle $, thus the wave function can be
derived as%
\begin{align}
\Psi_{H}\left(  p\right)   &  =\left \langle p\right \vert \left.
\Psi \right \rangle _{H}=N_{\mu,\nu}\left \langle p\right \vert SS^{\dag}%
H_{n}\left(  \hat{O}\right)  S\left \vert 0\right \rangle \nonumber \\
&  =\frac{N_{\mu,\nu}}{\sqrt{u}}\left \langle \frac{p}{u}\right \vert
H_{n}\left(  \hat{O}_{1}\right)  \left \vert 0\right \rangle . \label{h26}%
\end{align}
Now we calculate the matrix element $\left \langle p\right \vert H_{n}(\hat
{O})\left \vert 0\right \rangle $. Using the normal ordering form of $H_{n}%
(\hat{O})$ and Eq.(\ref{h6}), ($e^{\alpha a}a^{\dag}e^{-\alpha a}=a^{\dag
}+\alpha$), and noticing that \cite{27} $\left \vert p\right \rangle =\pi
^{-1/4}\exp \{-\frac{1}{2}p^{2}+\sqrt{2}ipa^{\dag}+\frac{1}{2}a^{\dag2}\}
\left \vert 0\right \rangle $, we have ($\lambda=\nu/\sqrt{1-2\mu \nu}$)%
\begin{align}
&  \left \langle p\right \vert H_{n}\left(  \hat{O}\right)  \left \vert
0\right \rangle \nonumber \\
&  =\frac{1}{\pi^{1/4}}\left(  \sqrt{1-2\lambda^{2}}\frac{\nu}{\lambda
}\right)  ^{n}e^{-\frac{1}{2}p^{2}}H_{n}\left(  \frac{-i\sqrt{2}\lambda
p}{\sqrt{1-2\lambda^{2}}}\right)  . \label{h27a}%
\end{align}
Thus the distribution of the quadrature $p$ is given by%
\begin{align}
\left \vert \Psi_{H}\left(  p\right)  \right \vert ^{2}  &  =\frac{N_{\mu,\nu
}^{2}}{\pi^{1/2}u}e^{-u^{-2}p^{2}}\left \vert 1-2\mu_{1}\nu_{1}-2\nu_{1}%
^{2}\right \vert ^{n}\nonumber \\
&  \times \left \vert H_{n}\left(  \frac{-i\sqrt{2}\nu_{1}p/u}{\sqrt{1-2\mu
_{1}\nu_{1}-2v_{1}^{2}}}\right)  \right \vert ^{2}, \label{h28}%
\end{align}
where $\left(  u=e^{r}\right)  $ and $\mu_{1},\nu_{1}$ are defined above.
Eq.(\ref{h28}) is just the Hermite-Gaussian function.

In order to clearly see the squeezing effect, we present the distribution in
Fig.4 where the distributions are plotted for different values of $n$ and
($\mu,\nu$). Different from the Gaussian distribution of squeezed state (n=0),
the HPS-SV has several different peak distributions with different $n(\neq0)$
values. In addition, the amplitude values of peaks affected by the parameters
$\mu,\nu$ (see Fig.4(b)).

\begin{figure}[ptb]
\label{Fig3} \centering \includegraphics[width=6cm]{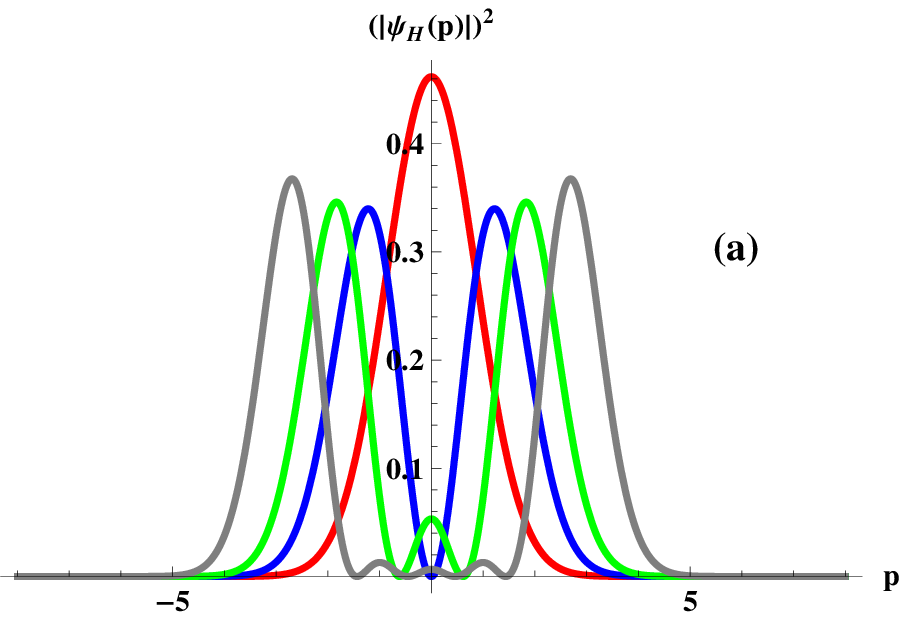} \newline%
\includegraphics[width=6cm]{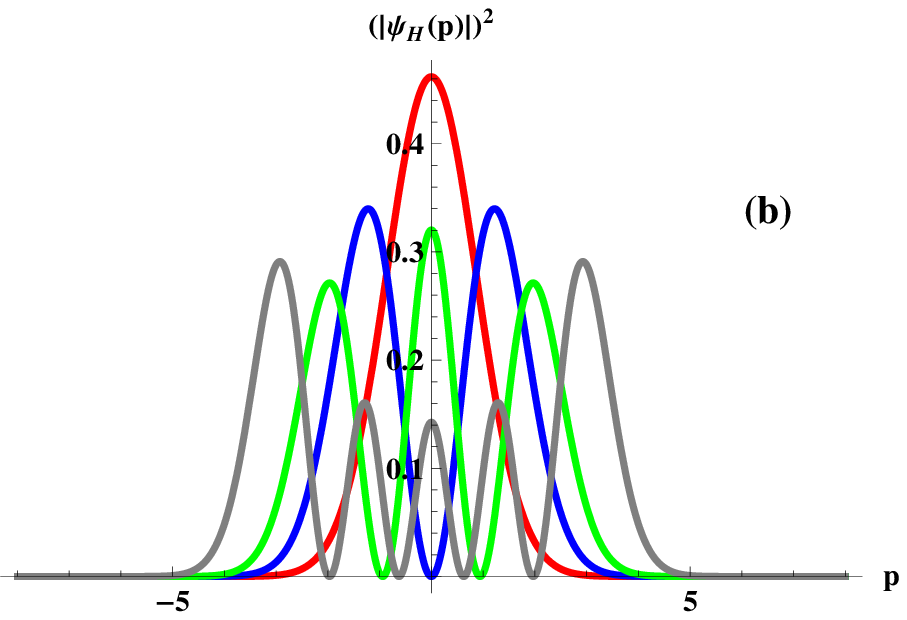}\caption{{}(Color online) The
distributions of the quadrature $p$ of the HPS-SV as a function of $p$ for
several different parameters $n$ and ($\mu,\nu$) with $r=0.2$. Red, Blue,
Green and Gray lines correspond to $n=0,1,2,4$, respectively. (a) $\mu
=\nu=1/\sqrt{2};$(b) n=1,2,4 correspond to $\left(  \mu,\nu \right)  =\left(
0,1\right)  $, $\left(  1,2\right)  $, $\left(  1,6\right)  $, respectively.}%
\end{figure}

Next, we further discuss the squeezing property of the HPS-SV by using the
standard analysis of quadrature squeezing, i.e., $\left(  \triangle Q\right)
^{2}<1$ or $\left(  \triangle P\right)  ^{2}<1$ which indicates the squeezing
or sub-Poissonian statistics. Here, we introduce a quadrature operator
$Q_{\theta}=ae^{-i\theta}+a^{\dagger}e^{i\theta}$. Thus the squeezing can be
characterized by the minimum value $\left \langle \triangle^{2}Q_{\theta
}\right \rangle $ $<1$ with respect to $\theta$, or by the normal ordering form
$\left \langle \colon \triangle^{2}Q_{\theta}\colon \right \rangle <0$ \cite{28}.
Upon expanding the terms of $\left \langle \colon \triangle^{2}Q_{\theta}%
\colon \right \rangle $, one can minimize its value over the whole angle
$\theta$, which is given by \cite{29} $S_{opt}=-2\left \vert \left \langle
a^{\dagger2}\right \rangle -\left \langle a^{\dagger}\right \rangle
^{2}\right \vert +2\left \langle a^{\dagger}a\right \rangle -2\left \vert
\left \langle a^{\dagger}\right \rangle \right \vert ^{2}$, then its negative
value in the range $\left[  -1,\text{ }0\right)  $ indicates squeezing (or
nonclassical). For the HPS-SV, $\left \langle a^{\dagger}\right \rangle =0,$
using Eq.(\ref{h16}), the degree of squeezing of the HPS-SV can be obtained%
\begin{equation}
S_{HPS}=2\left \{  \left \langle a^{\dagger}a\right \rangle _{H}-\left \vert
\left \langle a^{\dagger2}\right \rangle _{H}\right \vert \right \}  <0,
\label{h30}%
\end{equation}
which indicates that the negative value of $S_{HPS}$ only emerges when
$\left \langle a^{\dagger}a\right \rangle _{H}<\left \vert \left \langle
a^{\dagger2}\right \rangle _{H}\right \vert ,$ where $\left \langle
a^{2}\right \rangle _{H}=\left \langle a^{2}\cosh^{2}r+a^{\dag2}\sinh
^{2}r-a^{\dag}a\sinh2r\right \rangle _{\mu \rightarrow \mu_{1},\nu \rightarrow
\nu_{1}}$-$\frac{1}{2}\sinh2r$. In particular, when $n=0\ $(i.e., the case of
squeezed vacuum), $S_{HPS}=-2e^{-r}\sinh r,$ as expected.

\begin{figure}[ptb]
\label{Fig5} \centering \includegraphics[width=6cm]{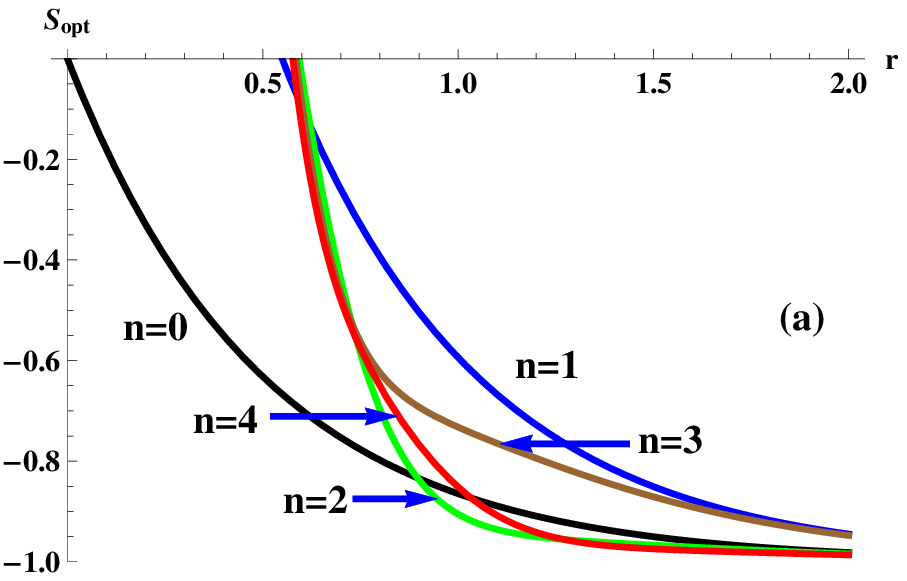} \newline%
\includegraphics[width=6cm]{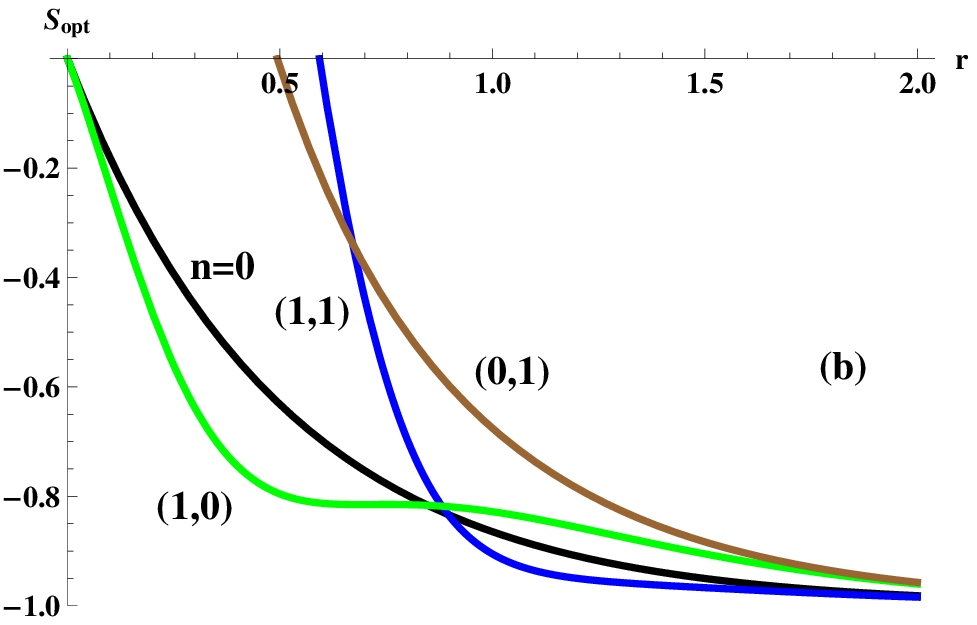}\  \caption{{}(Color online) The degree
of squeezing $S_{opt}$ of HPS-SV as the function of squeezing parameter $r$
for several different values of $n$ and ($\mu,\nu$). (a) ($\mu,\nu$)$=$%
($1,1$); (b) $n=2.$}%
\end{figure}

The degree of squeezing of the HPS-SV is shown in Fig. 5 for several different
values of parameters $n$ and ($\mu,\nu$). It is found that the degree of
squeezing of the HPS-SV increases with $r$. In Fig.5 (a) with a given
($\mu,\nu$)=(1,1), comparing with the SV (n=0), the HPS-SV can present
squeezing only when the squeezing parameter $r$ exceeds a certain threshold
value (say 0.5); the degree of squeezing can be enhanced by Hermite
polynomials superposition operation (say $n=2,4$) in a larger region of
squeezing parameter. In Fig.5 (b) with $n=2$, it is shown that (i) the Hermite
photon-subtraction operation $H_{2}\left(  a\right)  $ on the SV can be used
to improve the degree of squeezing in a small region ($r\lesssim0.6$), while
for $H_{2}\left(  a^{\dag}\right)  $ the case is not true; the coherent
superposition operation ($H_{2}\left(  a+a^{\dag}\right)  $), rather than the
$H_{2}\left(  a\right)  $ or $H_{2}\left(  a^{\dag}\right)  $ operations, can
improve the degree of squeezing in a large region. This implies that the
coherent operation $\mu a+\nu a^{\dag}$ achieves better squeezing than the
mere photon-subtraction (-addition) in a large region of $r$. In addition, the
maximum degree of squeezing of the HPS-SV is $-1$.

\section{Wigner distribution of the HPS-SV}

As a kind of quasi-probability function, the Wigner function (WF) is a
powerful tool to describe the nonclassicality of optical fields, whose partial
negativity implies the highly nonclassical properties of quantum states. In
addition, the negativity is often used to present the decoherence of quantum
states. In this section, we derive the analytical expression of WF for the
HPS-SV by using the the Weyl ordered operators' invariance under similar
transformations \cite{30}. For a single-mode quantum system, the WF can be
calculated as $W=$tr$\left[  \rho \Delta \left(  \alpha \right)  \right]  $,
where $\Delta \left(  \alpha \right)  $ single-mode Wigner operator
\cite{30,31},%
\begin{align}
\Delta \left(  \alpha \right)   &  =\frac{e^{2\left \vert \alpha \right \vert ^{2}%
}}{\pi}%
{\displaystyle \int}
\frac{d^{2}\beta}{\pi}\left \vert \beta \right \rangle \left \langle
-\beta \right \vert e^{2\left(  \alpha \beta^{\ast}-\alpha^{\ast}\beta \right)
}\nonumber \\
&  =\frac{1}{2}%
\genfrac{}{}{0pt}{}{\colon}{\colon}%
\delta \left(  \alpha-a\right)  \delta \left(  \alpha^{\ast}-a^{\dag}\right)
\genfrac{}{}{0pt}{}{\colon}{\colon}%
. \label{h33}%
\end{align}
Here $\alpha=\left(  q+ip\right)  /\sqrt{2}$ and the symbol $%
\genfrac{}{}{0pt}{}{\colon}{\colon}%
\genfrac{}{}{0pt}{}{\colon}{\colon}%
$ denotes Weyl ordering.

The merit of Weyl ordering lies in the Weyl ordered operators' invariance
under similar transformations proved, which means $S%
\genfrac{}{}{0pt}{}{:}{:}%
\left(  \circ \circ \circ \right)
\genfrac{}{}{0pt}{}{:}{:}%
S^{-1}=%
\genfrac{}{}{0pt}{}{:}{:}%
S\left(  \circ \circ \circ \right)  S^{-1}%
\genfrac{}{}{0pt}{}{:}{:}%
$, as if the \textquotedblleft fence" $%
\genfrac{}{}{0pt}{}{:}{:}%
\genfrac{}{}{0pt}{}{:}{:}%
$did not exist, so $S$ can pass through it. Then following this invariance and
the above squeezing transform relations, we have%
\begin{equation}
S^{\dagger}\Delta \left(  \alpha \right)  S=\frac{1}{2}%
\genfrac{}{}{0pt}{}{\colon}{\colon}%
\delta \left(  \bar{\alpha}-a\right)  \delta \left(  \bar{\alpha}^{\ast}%
-a^{\dag}\right)
\genfrac{}{}{0pt}{}{\colon}{\colon}%
=\Delta \left(  \bar{\alpha}\right)  , \label{h35}%
\end{equation}
where $\bar{\alpha}=\alpha \cosh r+\alpha^{\ast}\sinh r$. Thus the WF can be
derived
\begin{equation}
W\left(  \alpha,\alpha^{\ast}\right)  =N_{\mu_{1},\nu_{1}}^{2}\left \langle
0\right \vert H_{n}\left(  \hat{O}_{1}^{\dag}\right)  \Delta \left(  \bar
{\alpha}\right)  H_{n}\left(  \hat{O}_{1}\right)  \left \vert 0\right \rangle .
\label{h36}%
\end{equation}
Further employing Eq.(\ref{h6}) we can obtain
\begin{align}
W\left(  \alpha,\alpha^{\ast}\right)   &  =\frac{e^{-2\left \vert \bar{\alpha
}\right \vert ^{2}}}{\pi N_{\mu_{1},\nu_{1}}^{-2}}\frac{\partial^{2n}}%
{\partial \tau^{n}\partial t^{n}}\nonumber \\
&  \times \left.  e^{4\nu_{1}\bar{\alpha}\tau+4\nu_{1}\bar{\alpha}^{\ast
}t-\allowbreak4\nu_{1}^{2}t\tau+\left(  2\mu_{1}\nu_{1}-1\right)  \left(
\tau^{2}+t^{2}\right)  }\right \vert _{\tau=t=0}\nonumber \\
&  =\frac{1}{\pi}%
{\displaystyle \sum \limits_{l=0}^{n}}
\frac{\left(  n!\right)  ^{2}\left(  -\allowbreak4\nu_{1}^{2}\right)
^{l}\left(  2\mu_{1}\nu_{1}-1\right)  ^{n-l}}{l!\left[  \left(  n-l\right)
!\right]  ^{2}N_{\mu,\nu}^{-2}e^{2\left \vert \bar{\alpha}\right \vert ^{2}}%
}\nonumber \\
&  \times \left \vert H_{n-l}\left(  \frac{2\nu_{1}\bar{\alpha}}{i\sqrt{2\mu
_{1}\nu_{1}-1}}\right)  \right \vert ^{2}. \label{h37}%
\end{align}
Obviously, the WF $W\left(  \alpha,\alpha^{\ast}\right)  $ in Eq.(\ref{h37})
is a real function and is non-Gaussian in phase space due to the presence of
$H_{n-l}\left(  x\right)  $. In particular, when $n=0,$ Eq.(\ref{h37}) just
reduces to the WF of the squeezed vacuum, $\frac{1}{\pi}e^{-2\left \vert
\alpha \cosh r+\alpha^{\ast}\cosh r\right \vert ^{2}}$, as expected. In
addition, when $n=1$ corresponding to the single photon-subtraction
(-addition) squeezed vacuum, ($H_{0}\left(  x\right)  =1,H_{1}\left(
x\right)  =2x$)
\begin{equation}
W\left(  \alpha,\alpha^{\ast}\right)  =\frac{e^{-2\left \vert \bar{\alpha
}\right \vert ^{2}}}{\pi}\left(  4\left \vert \bar{\alpha}\right \vert
^{2}-\allowbreak1\right)  , \label{h38}%
\end{equation}
which indicates that there is always negative region at the center of phase
space $\alpha=0$ (independent of the two parameters $\mu,\nu$). In Fig.6, the
Wigner distributions are depicted in phase space for several different
parameter values $n$, and $\left(  \mu,\nu \right)  $, from which it clearly
see that there are some obvious negative regions of the WF in the phase space
which is an indicator of the nonclassicality of the state. In addition, these
negative areas are modulated not only by $n$ [see Fig.6 (a)-(c)], but also by
the parameters $\left(  \mu,\nu \right)  $ [see Fig.6 (b) and (d)]. For
instance, there is obvious difference of WF distribution between Fig.6 (b) and
(d). That is to say, for higher order case $n\geqslant2$, the negative area
depends on the two parameters. In order to clearly see this point, we can
quanlify the negative volume of the WF, defined by $\delta$=$\frac{1}{2}%
$[$\int_{-\infty}^{\infty}dqdp\left \vert W(q,p)\right \vert $-1] \cite{32}.

\begin{figure}[ptb]
\label{Fig6} \centering \includegraphics[width=8cm]{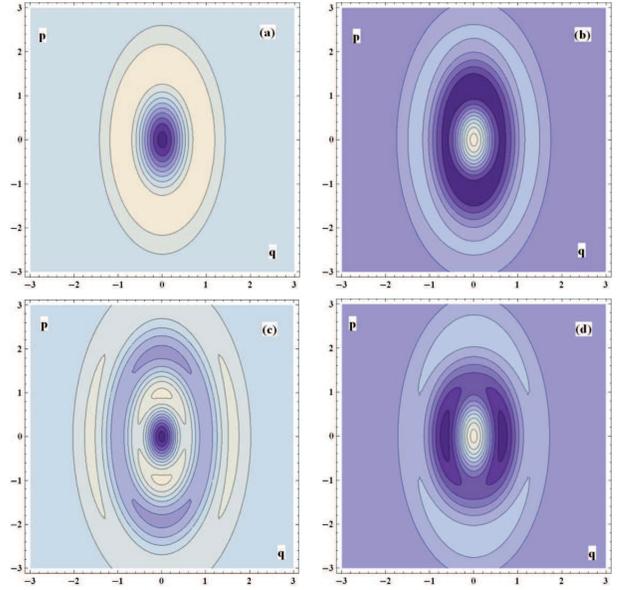}\caption{{}(Color
online) Contour plot of the Wigner function after the Hermite polynomials
coherent operation on the SV with $r=0.3$ for several different n and
($\mu,\nu$). (a) $n=1$; (b) $n=2$, $\mu,\nu=1;$ (c) $n=3$, $\mu,\nu=1;$ (d)
$n=2$, $\mu=1$, $\nu=9$, where only the negative regions are colored in blue.}%
\end{figure}

In Fig. 7, the negative volume of WF as a function of $r$ or $\nu$ on applying
the Hermite coherent suposition operatopn $H_{n}(\mu a+\nu a^{\dagger})$ for
several different $n$. From Fig.7(a), one can find that the negative volume
$\delta$ increases with the order $n$ (in a certain region of $r\lesssim0.45$)
and decreases with $r$. In particular, for the case of $n=1$ (corresponding to
a superposition between single-photon addition/subtraction SV and SV), the
negative volume is independent of parameters $r$ and ($\mu,\nu$)\ and is kept
unchanged ($\delta=0.2131$). In fact, using Eq.(\ref{h38}) one can calculate
that the negative volume of WF with $n=1$ is $\delta=2/\sqrt{e}-1\approx
0.2131$. In Fig.7(b), we optimize the negative volume for different $r$ and
$n,$ where $\mu$,$\nu$ are taken as $\mu=\sqrt{1-\nu^{2}}$. From Fig.7(b), it
is found that (i) when $\mu=\nu=1/\sqrt{2}$, the negative volume $\delta$
increases with the order $n$ for a given small $r$ and decreases with $r$ for
a given $n$ (see the vertical dotted line at the point of $\nu=1/\sqrt{2}$);
(ii) for a given parameter $r=0.1$, the negative volume $\delta$ increases
with $n$ when $\nu$ exceeds a certain threshold ($\nu \approx0.41$); (iii) for
a given $n=2,$ $\delta$ decreases with $r$ when $\nu$ exceeds a certain
threshold ($\nu \approx0.45$); (iv) the negative volume $\delta$ does not
monotonously increase with $\nu;$ in particular, one can find the maximum
negative volume $\delta$ in a bigger region of $\nu$ ($\nu \geqslant0.41$) for
$n=2,3,4$. This optimal value of $\delta$ can be achieved at neither $\nu=0$
nor $\nu=1$. For instance, this points are $\nu \approx0.71,0.74,0.78$ for
different values of $n=2,3,4$ and $r=0.1$. These indicate that the effects of
the coherent operation $H_{n}(\mu a+\nu a^{\dagger})$ with higher order
$n\geqslant2$ are prominent than those of the mere photon subtraction
$H_{n}(\nu a^{\dagger})$ and the addition $H_{n}(\nu a^{\dagger})$
particularly in the larger region of parameter $\nu$, whereas the optimal
operation is not the photon subtraction or the photon addition in this region.
This result is different from that in Ref.\cite{11}.

\begin{figure}[ptb]
\label{Fig7} \centering \includegraphics[width=6cm]{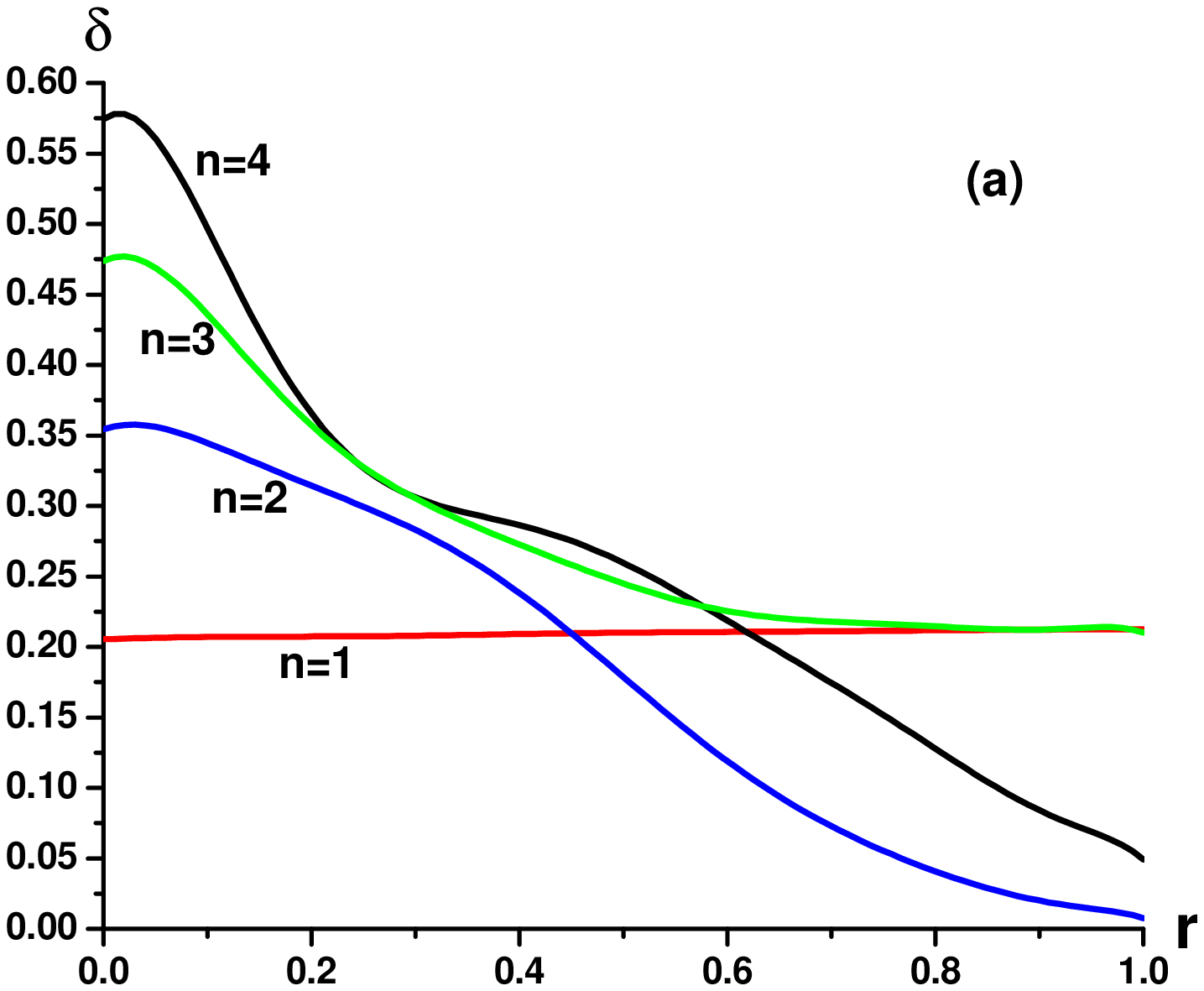}\  \newline%
\includegraphics[width=6cm]{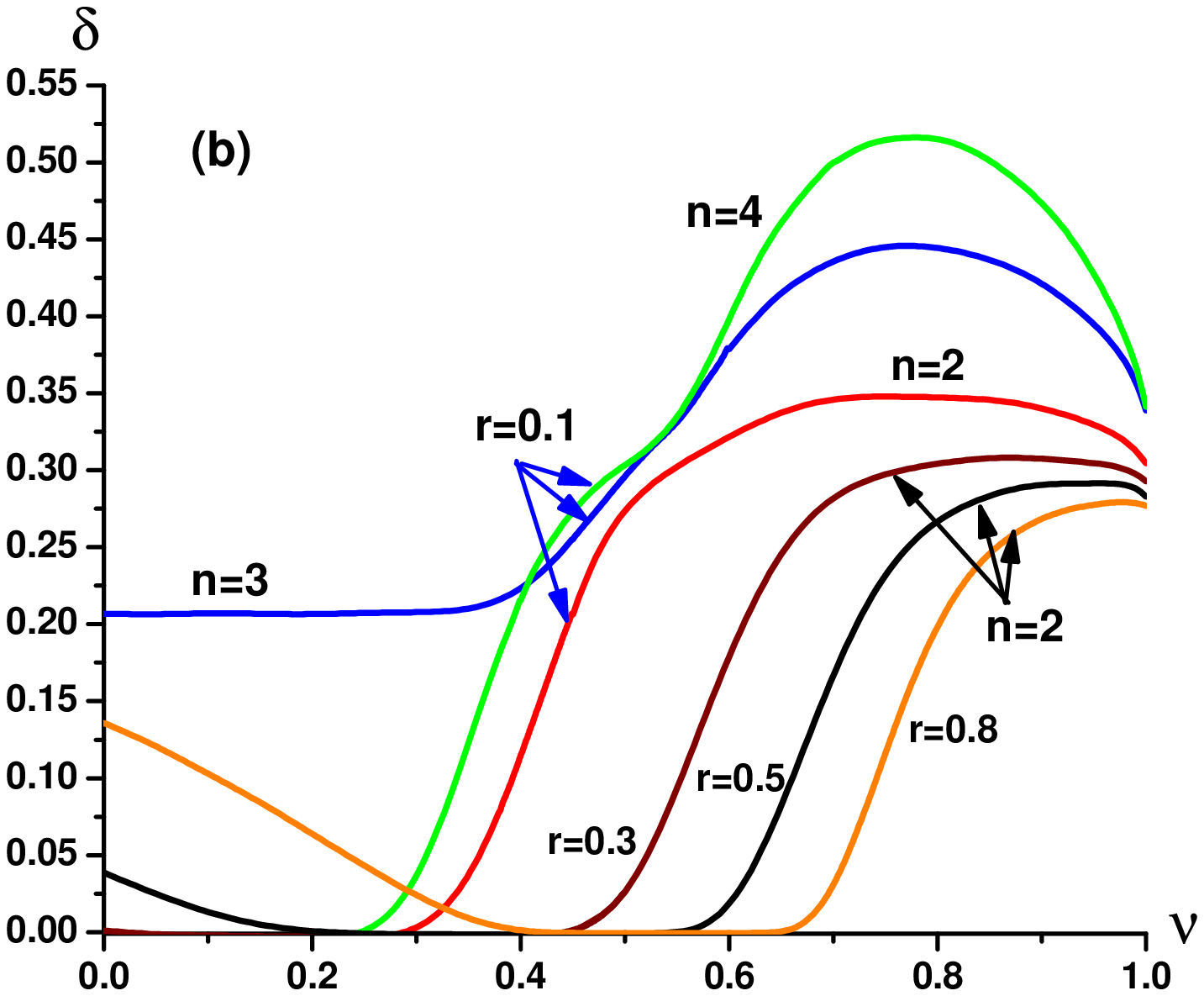}\caption{{}(Color online) Negative
volume of the Wigner function as a function of (a) squeezing parameter $r$,
and $\mu$=$\nu$=$1/\sqrt{2}$; (b) parameter $\nu$ with $\mu=\sqrt{1-\nu^{2}}$
for several different $n=1,2,3,4$, and $r=0.1,0.3,0.5,0.8.$}%
\end{figure}

\section{Decoherence of the HPS-SV in Phase-Sensitive Reservoirs}

In this section, we shall examine the time evolution of the HPS-SV at the
presence of phase-sensitive reservoirs. In the interaction picture and the
Born and Markow approximation, the time evolution of the density matrix is
governed by the master equation (ME) \cite{33}:%
\begin{align}
\frac{d}{dt}\rho \left(  t\right)   &  =\kappa \bar{n}L\left[  a^{\dag}\right]
\rho+\kappa \left(  \bar{n}+1\right)  L\left[  a\right]  \rho \nonumber \\
&  +\kappa MD\left[  a\right]  \rho+\kappa M^{\ast}D\left[  a^{\dag}\right]
\rho, \label{t1}%
\end{align}
and%
\begin{align}
L\left[  O^{\dag}\right]  \rho &  =2O^{\dag}\rho O-OO^{\dag}\rho-\rho
OO^{\dag},\label{t2}\\
D\left[  a\right]  \rho &  =2a^{\dag}\rho a^{\dag}-a^{\dag2}\rho-\rho
a^{\dag2}. \label{t3}%
\end{align}
where $\kappa$ and $\bar{n}$ are the dissipative coefficient and the average
thermal photon number of the environment, respectively. Here $M$ is the
complex correlation parameter between modes symmetrically displaced about
center frequency. In fact, the ME in Eq.(\ref{t1}) includes two special cases:
(1) For an uncorrelated reservoir, i.e., $M=0,$ Eq.(\ref{t1}) becomes the ME
describing the interaction between a system and a thermal environment at
finite temperature; (2) when $M=\bar{n}=0,$ Eq.(\ref{t1}) reducess to the one
describing the photon-loss channel. For an (non)ideally squeezed reservoir,
the constraint condition $\left \vert M\right \vert ^{2}=\bar{n}\left(  \bar
{n}+1\right)  $ ($\left \vert M\right \vert ^{2}<\bar{n}\left(  \bar
{n}+1\right)  $) is required.

In Ref.\cite{34}, we derived the Kraus operator-sum representation of density
operator $\rho$ and the time evolution of some distibution functions by using
the thermal entangled state representation $\left \langle \eta \right \vert $.
The evolution of Wigner function is given by%
\begin{equation}
W\left(  \alpha,t\right)  =\frac{2\mu_{\infty}}{T}\int \frac{d^{2}\beta}{\pi
}e^{-\frac{2\mu_{\infty}^{2}}{T}\Sigma \left(  \bar{\alpha},\bar{\alpha}^{\ast
}\right)  }W\left(  \beta,0\right)  , \label{t4}%
\end{equation}
where $T=1-e^{-2\allowbreak \kappa t},$ $\bar{\alpha}=\alpha-\beta e^{-\kappa
t}$, and $\mu_{\infty}=1/\sqrt{\left(  2\bar{n}+1\right)  ^{2}-4\left \vert
M\right \vert ^{2}}$, $\Sigma \left(  \bar{\alpha},\bar{\alpha}^{\ast}\right)  $
is defined as%
\begin{align}
\Sigma \left(  \bar{\alpha},\bar{\alpha}^{\ast}\right)   &  =\left(
\begin{array}
[c]{cc}%
\bar{\alpha} & \bar{\alpha}^{\ast}%
\end{array}
\right)  \sigma_{\infty}\left(
\begin{array}
[c]{c}%
\bar{\alpha}\\
\bar{\alpha}^{\ast}%
\end{array}
\right)  ,\label{t5}\\
\sigma_{\infty}  &  =\left(
\begin{array}
[c]{cc}%
M^{\ast} & \bar{n}+\frac{1}{2}\\
\bar{n}+\frac{1}{2} & M
\end{array}
\right)  .\nonumber
\end{align}
Noting the differential expression of Wigner function Eq.(\ref{h37}), we can
finally obtain%

\begin{equation}
W\left(  \alpha,t\right)  =W_{r}\left(  \alpha,t\right)  F_{n}\left(
\alpha,t\right)  , \label{t7}%
\end{equation}
where $W_{r}\left(  \alpha,t\right)  $ is the evolution of Wigner function of
squeezed vacuum in phase sensitive resevoire, and $F_{n}\left(  \alpha
,t\right)  $ is a non-Gaussian item due to the presence of Hermite
excitation,
\begin{align}
W_{r}\left(  \alpha,t\right)   &  =\frac{\mu_{\infty}e^{-P}}{\pi \sqrt{D}%
}e^{\frac{1}{TD}\left(  2R_{1}R_{2}R_{2}^{\ast}-R_{3}R_{2}^{\ast}{}^{2}%
-R_{2}^{2}R_{3}^{\ast}\right)  },\label{t8}\\
F_{n}\left(  \alpha,t\right)   &  =\sum_{l=0}^{n}\frac{\left[  n!\right]
^{2}\left(  -G_{1}\right)  ^{l}\left \vert G_{2}\right \vert ^{n-l}}{l!\left[
\left(  n-l\right)  !\right]  ^{2}}\nonumber \\
&  \times N_{\mu,\nu}^{2}\left \vert H_{n-l}\left(  \frac{G_{3}}{2i\sqrt{G_{2}%
}}\right)  \right \vert ^{2}, \label{t9}%
\end{align}
and $D=R_{1}^{2}-\left \vert R_{3}\right \vert ^{2},$%
\begin{align}
P  &  =\frac{2\mu_{\infty}^{2}}{T}\left[  \left(  2\bar{n}+1\right)
\left \vert \alpha \right \vert ^{2}+M\alpha^{\ast}{}^{2}+M^{\ast}\alpha
^{2}\right]  ,\nonumber \\
R_{1}  &  =\left(  1+2\bar{n}\right)  \mu_{\infty}^{2}e^{-2(t\kappa
)}+\allowbreak T\allowbreak \cosh2r,\nonumber \\
R_{2}  &  =\mu_{\infty}^{2}\left(  \alpha^{\ast}+2\bar{n}\alpha^{\ast}+2\alpha
M^{\ast}\right)  e^{-\kappa t},\nonumber \\
R_{3}  &  =2\mu_{\infty}^{2}M^{\ast}e^{-2(t\kappa)}+T\sinh2r, \label{t10}%
\end{align}
as well as%
\begin{align}
G_{1}  &  =4\nu_{1}^{2}\left \{  1+\frac{T}{D}\left[  \left(  R_{3}+R_{3}%
^{\ast}\right)  \sinh2r-2R_{1}\cosh2r\right]  \right \}  ,\nonumber \\
G_{2}  &  =2\mu_{1}\nu_{1}-1+\frac{2T\nu_{1}^{2}}{D}\left \{  2R_{1}%
\sinh2r\right. \nonumber \\
&  \left.  -\allowbreak \left(  R_{3}-R_{3}^{\ast}+\left(  R_{3}+R_{3}^{\ast
}\right)  \cosh2r\right)  \right \}  ,\nonumber \\
G_{3}  &  =\frac{4\nu_{1}}{D}\allowbreak R_{1}\left(  \allowbreak R_{2}^{\ast
}\sinh r+R_{2}\cosh r\allowbreak \right) \nonumber \\
&  -\frac{4\nu_{1}}{D}\allowbreak \left(  R_{2}^{\ast}R_{3}\cosh r+R_{3}^{\ast
}\allowbreak R_{2}\sinh r\right)  . \label{t11}%
\end{align}

In partciular, at the center of phase space $\alpha=0$, we have $R_{2}%
=0,P=0,G_{3}=0.$ Thus for the case of $n=1$, we can get $W\left(
\alpha,t\right)  \propto-G_{1}$. Thus the existence of negative volume of WF
is determined by $G_{1}>0,$ which leads to
\begin{equation}
\kappa t<\kappa t_{c}=\frac{1}{2}\ln \left(  \mu_{\infty}+1\right)  ,
\label{t12}%
\end{equation}
which is independent of squeezing parameter $r$. It is easy to see that for
any $M$ ranging from $0$ to $\bar{n}(\bar{n}+1)$,
\begin{equation}
\frac{1}{2}\ln \left(  \frac{2\bar{n}+2}{2\bar{n}+1}\right)  \leqslant \frac
{1}{2}\ln \left(  \mu_{\infty}+1\right)  \leqslant \frac{1}{2}\ln2, \label{t13}%
\end{equation}
which indicates that the characteristic time of decoherence of single-photon
added squeezed vacuum state in phase sensitive reservoirs is larger than that
in the thermal enviornment and smaller than that in photon-loss channel.

\begin{figure}[ptb]
\label{Fig8} \centering \includegraphics[width=8cm]{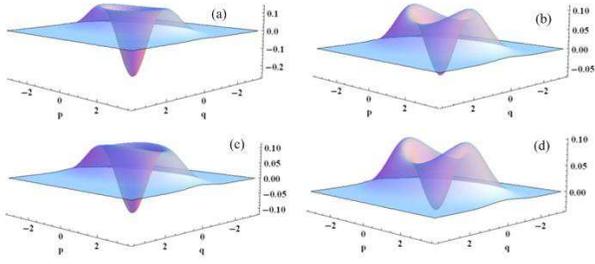}\caption{{}(Color
online) The evolution of Wigner function distribution in phase space with
$n=1,$ $r=0.3$. (a) $\kappa t=0.01,$ $M=0.1,$ $\bar{n}=1;$ (b) $\kappa
t=0.08,$ $M=0.1,$ $\bar{n}=1;$ (c) $\kappa t=0.08,$ $M=1,$ $\bar{n}=1;$ (d)
$\kappa t=0.08,$ $M=0.1,$ $\bar{n}=1.5;$}%
\end{figure}

In order to measure the degree of nonclassicality for the evoluted state, we
consider the negative area and the nagative volume in phase space. As shown in
Fig. 8, it is shown that the nagative area gradually dissapears with the
increasemenet of $\bar{n}$, $\kappa t$, while increases with parameter $M$. To
clearly see the effects of the decoherence and parameter $M$ on the
nonclassical properties, the evolutions of negative volume with time and
squeezing parameter $M$ are plotted in Fig.9 for given $n=1$, $r=0.3$ and
$\bar{n}=0.5$. From Fig.9(a) one can see that the nagative volume
monotonically diminnishes with $\kappa t$, and there is a more rapid
attenuation for a big $n$ than a small one; this leads to a smaller negative
volume for a big $n$ than a small one when $\kappa t$ ecseeds a certain value.
From this point, one can draw a conclusion that single-photon
subtraction/addition SV present a more strongger roboustness against the
reservoirs than a higher-order photon subtraction/addition which could have a
bigger negative volume at initial time. From Fig.9(b), it is shown that the
parameter $M$ can be used to enhance the nonclassicality of quantum state in a
phase-sensitive reservoirs. Specially speaking, the negative volume increases
monotonically with $M$. The optimal volume appears at the the maximum value of
$\left \vert M\right \vert ^{2}=\bar{n}(\bar{n}+1)$, which inreases with $n$, as expected.

\begin{figure}[ptb]
\label{Fig9} \centering \includegraphics[width=6cm]{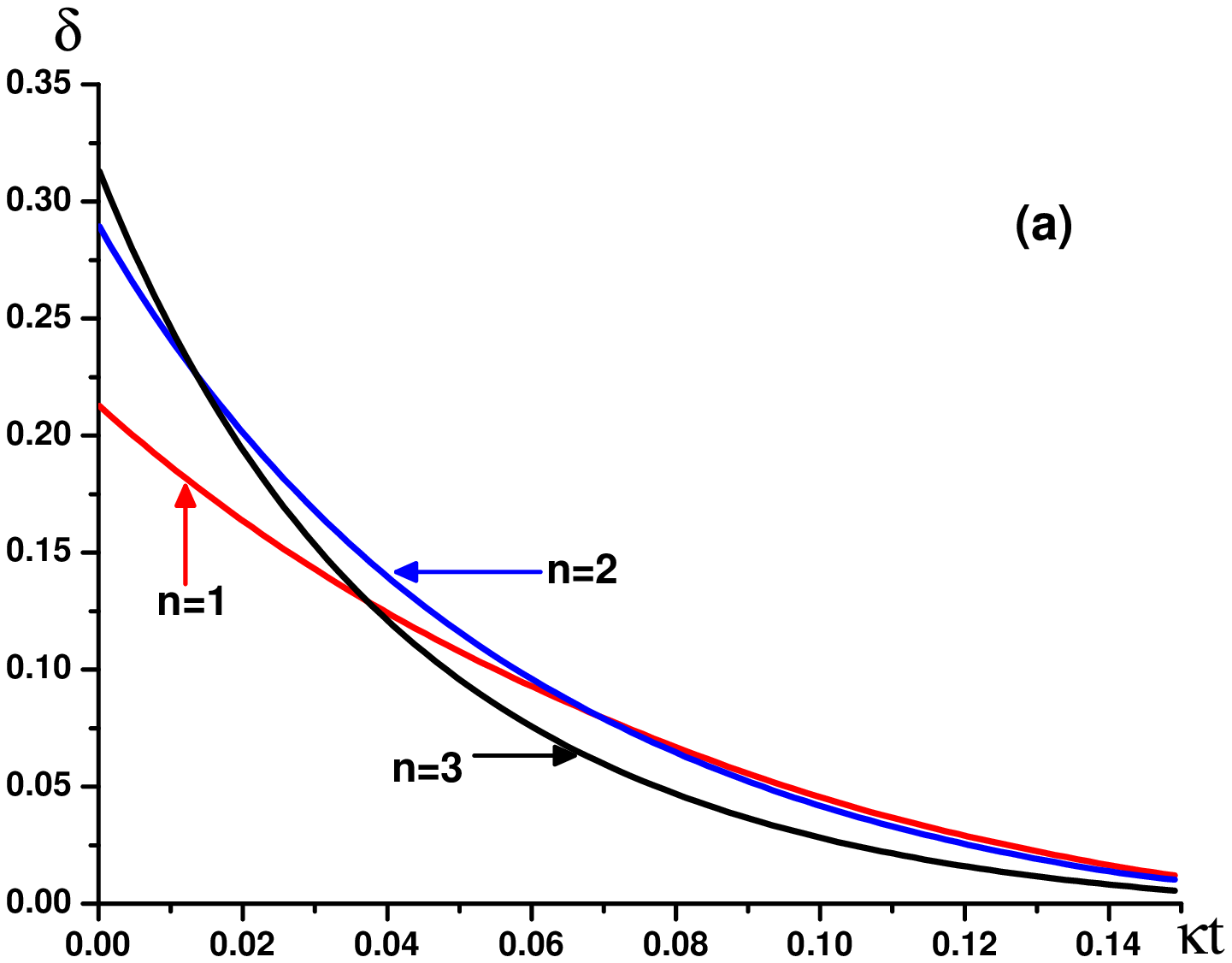}\  \newline%
\includegraphics[width=6cm]{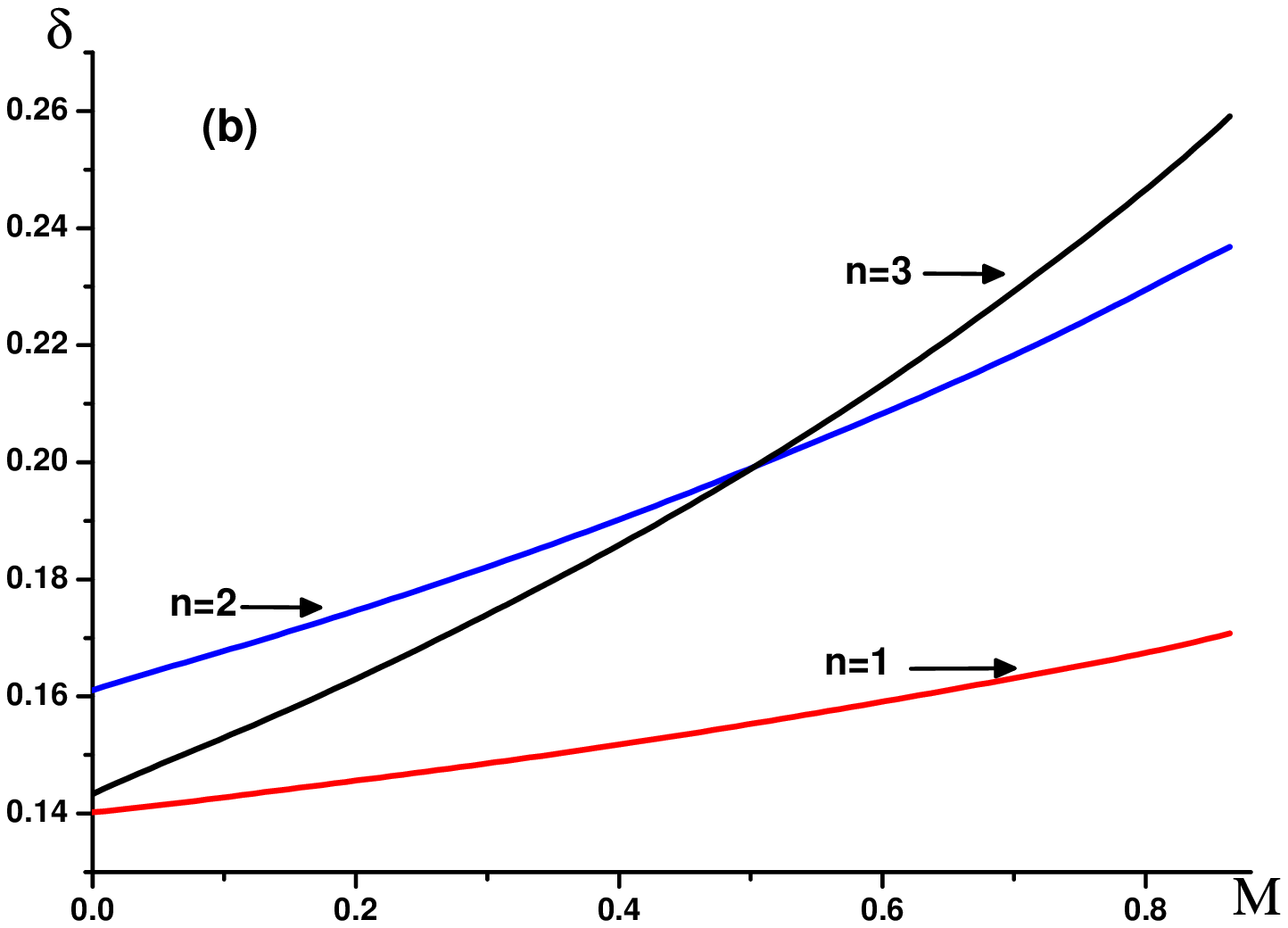}\caption{{}(Color online) Evolution of
Negative volume of Wigner function as a function of (a) $\kappa t$ with
$M=0.1$; (b) $M$ with $\kappa t=0.03$ for several different $n=1,2,3$, and
$r=0.3$, $\bar{n}=0.5,$ as well as $\mu$=$\nu$=$1/\sqrt{2}$.}%
\end{figure}

\section{Conclusions}

In this paper, we have introduced a new non-Gaussian state, which is generated
by applying Hermite-polynomial excitation on squeezed vacuum. Single-photon
addition/subtraction and Hermite polynomial's addition/subtraction can be seen
as its three special cases. Its normalized factor is found to be a Legendre
polynomial. Then we have investigated its nonclassicality according to the
Mandel's Q parameter, second-order correlation function, photon-number
distribution, squeezing effects and the negativity of WF in phase space. It is
shown that all these nonclassical properties can be obviously improved by the
HPS operation and can be remarkably modulated by superposition parameters
$\mu$ and $\nu$. The degree of squeezing of the HPS-SV increases with $r$. In
particular, comparing with the SV, an observable improvement of squeezing
effect can be achieved by Hermite photon-subtraction operation $H_{2}\left(
a\right)  $ and the coherent superposition operation ($H_{2}\left(  a+a^{\dag
}\right)  $) in a small region ($r\lesssim0.6$) and a large region, respectively.

In addition, the numerical calculation of negative volume $\delta$ of WF
showed that $\delta$ increases with the order $n$ for $r\lesssim0.45$ and
decreases with $r$. It is interesting to notice that the negative volume
$\delta$ is given $\delta=2/\sqrt{e}-1\approx0.2131$ independent of $\mu$ and
$\nu$, as well as $r$. For high-order excitation ($n\geqslant2$), the negative
volume can be optimized by modulating parameters $\nu$ ($\mu=\sqrt{1-\nu^{2}}%
$) and $n$ for a given $r$. It is found that the negative volume $\delta$ may
increase with $n$ and decrease with $r$ when $\nu$ exceeds a certain
threshold. In particular, the optimal value of $\delta$ can be obtained in a
bigger region of $\nu$ ($1>\nu \gtrsim0.41$)\ not at $\nu=0$ or $\nu=1$. This
implies that the effects of the coherent operation $H_{n}(\mu a+\nu
a^{\dagger})$ with higher order $n\geqslant2$ are prominent than those of the
mere photon subtraction $H_{n}(\nu a^{\dagger})$ and the addition $H_{n}(\nu
a^{\dagger})$ particularly in the larger region of parameter $\nu$, whereas
the optimal operation is not the photon subtraction or the photon addition in
this region, which is a new result.

Furthermore, we have considered the decoherence effects of the HPS-SV in
phase-sensitive reservoirs according to the analitically derived WF
distribution. It is shown that the negative area and volume diminish gradually
with the evolution of time and dissapear eventually. However, the negative
volume of the HPS-SV with higher order excitation decays more rapidly with
time, which implies that single-photon subtraction/addition SV has a more
strongger roboustness than a higher-order photon subtraction/addition although
the latter has a bigger negative volume at initial time. In addition, the
parameter $M$ describing the squeezing characteristic of reservoirs can be
effectively used to enhance the nonclassicality.

\textbf{Acknowledgments: }Project supported by the National Natural Science
Foundation of China (Grant No.11264018), the Natural Science Foundation of
Jiangxi Province of China (Grant No. 20132BAB212006), and the Research
Foundation of the Education Department of Jiangxi Province of China (no
GJJ14274) as well as Degree and postgraduate education teaching reform project
of jiangxi province(No. JXYJG-2013-027).

\end{document}